\documentclass[aps,10pt,twocolumn,prl,amsmath,amssymb,superscriptaddress,longbibliography]{revtex4-2}
\usepackage[T1]{fontenc}
\usepackage{amsmath,amsfonts,amssymb,mathtools,upgreek}
\usepackage{hyperref}
\hypersetup{hidelinks,pdftitle={Full macroscopic thermalization and the formation of Schrodinger's cats in the weakly perturbed Ising model},pdfauthor={Hal Tasaki}}
\def\eq#1\en{\begin{equation}#1\end{equation}}
\def\eqa#1\ena{\begin{align}#1\end{align}}
\def\eqg#1\eng{\begin{gather}#1\end{gather}}
\newcommand{\lb}[1]{\label{e:#1}}
\newcommand{\rlb}[1]{\eqref{e:#1}}
\newcommand{\nl}{\notag\\}

\newcommand{\para}[1]{\medskip\par{\em #1}\/.---}
\newcommand{\prop}[1]{{\em #1}\/.---}
\newcounter{ct}
\newcommand{\ctl}[1]{\refstepcounter{ct}\thect\label{#1}}
\newcommand{\qedm}{\hfill$\square$}

\newcommand{\norm}[1]{\left\Vert#1\right\Vert}
\newcommand{\bkt}[1]{\left\langle#1\right\rangle}
\newcommand{\ket}[1]{|#1\rangle}
\newcommand{\bra}[1]{\langle#1|}
\newcommand{\Tr}{\operatorname{Tr}}
\newcommand{\Proj}{\operatorname{Proj}}

\newcommand{\osc}{\operatorname{osc}}
\newcommand{\Prob}{\operatorname{Prob}}
\newcommand{\E}{\mathbb E}
\newcommand{\Zb}{\mathbb Z}
\newcommand{\Cb}{\mathbb C}
\newcommand{\bc}{\mathrm{bc}}
\newcommand{\pp}{\mathrm p}
\newcommand{\can}{\mathrm{can}}
\newcommand{\mc}{\mathrm{mc}}
\newcommand{\neqs}{\mathrm{neq}}
\newcommand{\out}{\mathrm{out}}
\newcommand{\ms}{m_{\mathrm s}}
\newcommand{\betac}{\beta_{\mathrm c}}
\newcommand{\calA}{{\cal A}}

\newcommand{\calH}{{\cal H}}
\newcommand{\calO}{{\cal O}}
\newcommand{\calS}{{\cal S}}
\newcommand{\calU}{{\cal U}}

\newcommand{\ttH}{\widetilde H}
\newcommand{\ttcalH}{\widetilde{\calH}}
\newcommand{\ttP}{\widetilde P}
\newcommand{\ttOmega}{\widetilde\Omega}
\newcommand{\bssigma}{\boldsymbol{\sigma}}
\newcommand{\bseta}{\boldsymbol{\eta}}
\newcommand{\epsL}{\varepsilon_L}
\newcommand{\CR}{C_{\mathrm R}}

\begin{document}
\title{\texorpdfstring{Full macroscopic thermalization and the formation of Schr\"odinger's cats\protect\\
by unitary time evolution in the weakly perturbed Ising model\protect\\
--- Applications of the Roos--Sugimoto--Teufel--Tumulka--Vogel theory}{Full macroscopic thermalization and the formation of Schrodinger's cats by unitary time evolution in the weakly perturbed Ising model - Applications of the Roos-Sugimoto-Teufel-Tumulka-Vogel theory}}
\author{Hal Tasaki}
\email{hal.tasaki@gakushuin.ac.jp, hal.tasaki@gmail.com}
\affiliation{Department of Physics, Gakushuin University, Mejiro, Toshima-ku, Tokyo 171-8588, Japan}
\date{September 15, 2026}
\begin{abstract}
We study macroscopic thermalization in the ferromagnetic Ising model with a weak generic (highly nonlocal and many-body) random quantum perturbation. We prove that a single typical perturbation makes every initial state $\ket{\Phi(0)}$ in a specified energy shell thermalize: a measurement of a large class of macroscopic observables in the time-evolved state $\ket{\Phi(t)}=e^{-i\widetilde H_Lt}\ket{\Phi(0)}$ yields their thermal equilibrium values, within a prescribed precision and with overwhelming probability, at sufficiently large, typical times $t$.
The result covers the full finite-temperature range, including the ordered phase, where we use plus boundary conditions.
With periodic boundary conditions in the ordered phase, every energy eigenstate in the shell is shown to be an almost balanced macroscopic Schr\"odinger-cat state.
Moreover, every pure initial state becomes such a cat at sufficiently large, typical times, with each branch in the corresponding macroscopic thermal equilibrium. The proof combines the theory of Roos, Sugimoto, Teufel, Tumulka, and Vogel with rigorous Ising large-deviation estimates.
\end{abstract}
\maketitle

The minus-first law of thermodynamics asserts that an isolated macroscopic system, left undisturbed, eventually approaches thermal equilibrium \cite{BrownUffink}. An essential, logically distinct premise---a ``minus second law''?---is that, in thermal equilibrium, each macroscopic observable has a definite value determined by a small number of quantities, such as the total energy, volume, and amount of substance, together with any required specification of the thermodynamic phase. Equilibrium statistical mechanics represents these definite values as ensemble averages. This sets a goal for a quantum-mechanical foundation of equilibrium statistical mechanics: establish, for a suitable class of macroscopic quantum systems, that \emph{every} initial state consistent with the specified macroscopic constraints evolves unitarily into a state in which a one-shot measurement yields the corresponding thermal equilibrium value, with overwhelming probability at typical sufficiently late times. To recover the thermodynamic premise, the conclusion must concern an individual measurement outcome, not merely an average over times or repeated preparations.

In the present Letter, based on the recent theory of Roos, Sugimoto, Teufel, Tumulka, and Vogel \cite{Roos}, we give a rigorous example of this strong thermalization statement without relying on unproved assumptions.
The model is the standard ferromagnetic Ising model with a small quantum perturbation, and our theory covers the full finite-temperature range, notably including the low-temperature phase in $d\ge2$, where the model exhibits spontaneous symmetry breaking. As a rather striking additional conclusion, we prove that Schr\"odinger-cat states spontaneously emerge from unitary time evolution in the low-temperature phase. The price is a random, generally unphysical perturbation.

The macroscopic viewpoint on the quantum-mechanical foundations of statistical mechanics dates back to the seminal work of von Neumann and was further developed in subsequent works \cite{vN,GLMTZ,GLTZ,Reimann,Tasaki2016,Roos}. A central observation is that a suitable macroscopic version of the energy eigenstate thermalization hypothesis (ETH) implies equilibration or thermalization for every initial state in an energy shell. (Here, ``equilibration'' means relaxation to equilibrium; we use ``thermalization'' when the target is thermal equilibrium.) See also \cite{DAlessio,GE} for alternative approaches and broader perspectives.

In constructive approaches, one attempts to establish equilibration or thermalization for specified Hamiltonians without assuming ETH or other unproved properties \cite{Roos,Gluza,STvac,TasakiHeat,TasakiFree,HaraKoike}. Spatial equilibration from \emph{every} initial state of a specified nonrandom many-body system was first proved for a free fermion chain in \cite{TasakiFree}. Its coarse-grained particle density becomes uniform, while the momentum occupations remain conserved. Thus this is equilibration, not thermalization in the terminology used here.

Along with a careful examination of these free-fermion results, Roos, Sugimoto, Teufel, Tumulka, and Vogel \cite{Roos} developed a general theory treating highly degenerate spectra and weak generic random perturbations. Their free-fermion applications establish spatial-density equilibration. Our Ising application \cite{Earlier} addressed the thermal value of the nonzero spontaneous magnetization, determined by genuine many-body interactions.

Here we substantially extend the approach in \cite{Earlier} to study the Ising model with a small random perturbation and prove thermalization for \emph{every} initial state in a specified energy shell, simultaneously for all macroscopic observables in an increasing family. The theorem covers the disordered phase, including criticality, and the symmetry-broken plus phase. Periodic boundary conditions in the ordered phase give a different, rather striking conclusion: every pure initial state becomes an almost balanced superposition of the two macroscopic phases at typical late times. In other words, a Schr\"odinger-cat state is inevitably generated by unitary time evolution.

We must note, however, that the perturbation is unphysical; it is drawn from the full operator space and generally contains interactions of macroscopic range and order. Its admissible strength may depend severely on the size. One may hope that such a random perturbation captures certain aspects of realistic short-range perturbations; we return to this point in the Discussion.

\para{The Roos--Sugimoto--Teufel--Tumulka--Vogel theory}
Suppose an unperturbed Hamiltonian---the classical Ising Hamiltonian in our application---has a highly degenerate energy eigenspace admitting an orthonormal basis in which every basis state has negligible probability of a nonequilibrium measurement.
This is not ETH: degeneracy may allow other eigenstates that are far from equilibrium.
The theory in \cite{Roos} shows that the model with a typical sufficiently weak unitarily invariant random perturbation has every energy eigenstate in the corresponding shell in equilibrium, with a controlled, slightly larger error.
This is a version of ETH.

Let $P_{\neqs}$ be the projection onto the nonequilibrium subspace, and fix a perturbation with the above equilibrium property.
Since every normalized perturbed energy eigenstate $\ket\Psi$ in the shell obeys $\bra\Psi P_{\neqs}\ket\Psi\le\varepsilon^2/2$ for some small $\varepsilon>0$, the time average of $\bra{\Phi(t)}P_{\neqs}\ket{\Phi(t)}$ obeys the same bound for every initial state $\ket{\Phi(0)}$ in the energy shell.
This means that, for sufficiently large $T$, the set $\calA\subset[0,T]$ on which $\bra{\Phi(t)}P_{\neqs}\ket{\Phi(t)}>\varepsilon$ has length at most $\varepsilon T$, and hence occupies only a small fraction of the interval.
This proves macroscopic equilibration or thermalization at sufficiently large, typical times.

\para{The Ising model}
Let $\Lambda_L=\{1,\ldots,L\}^d$, $N=L^d$, and $L\ge3$.
The Hilbert space is $\calH_{\rm tot}=\bigotimes_{u\in\Lambda_L}\Cb^2$.
Write $X_u,Y_u,Z_u$ for the standard Pauli operators, with $Z_u\ket{\bssigma}=\sigma_u\ket{\bssigma}$ for classical configurations $\bssigma=(\sigma_u)_{u\in\Lambda_L}\in\{-1,1\}^{\Lambda_L}$.
We consider the Hamiltonians with periodic and plus boundary conditions:
\eqa
H_L^{\pp}&=-\sum_{u\in\Lambda_L}\sum_{j=1}^d Z_uZ_{u+e_j},\lb{Hp}\\
H_L^+&=-\sum_{\substack{\{u,v\}\subset\Lambda_L\\ |u-v|=1}}Z_uZ_v-\sum_{u\in\Lambda_L}n_u Z_u.\lb{Hplus}
\ena
Addition in \rlb{Hp} is modulo $L$; $e_j$ is the $j$th unit vector, and $n_u$ counts neighbors of $u$ outside the cube. We write $H_L^{\bc}$, $\bc=\pp,+$.
The canonical expectation $\bkt{\cdot}_{L,\beta}^{\can,\bc}$ is the normalized trace with weight $e^{-\beta H_L^{\bc}}$.

The magnetization density $M_L=N^{-1}\sum_u Z_u$ satisfies
\eq
\lim_{L\uparrow\infty}\bkt{M_L}_{L,\beta}^{\can,+}
=\ms(\beta)
\begin{cases}
=0,&\beta\le\betac(d),\\
{}>0,&\beta>\betac(d).
\end{cases}\lb{ms}
\en
Here $\betac(1)=\infty$, while $0<\betac(d)<\infty$ for $d\ge2$, with $\betac(2)=\frac12\log(1+\sqrt2)$.
The infinite-volume Gibbs state is unique for $\beta\le\betac(d)$; above $\betac(d)$ the plus boundary condition selects the plus state \cite{FV,ADS,DST}.
Throughout, $0<\beta<\infty$ is fixed before $L\uparrow\infty$.

\para{Energy shells and nonequilibrium projection}
Let $D_L^{\bc}(E)$ be the degeneracy of the energy eigenvalue $E$, and let $E_L^{\bc}(\beta)$ be an energy that maximizes $D_L^{\bc}(E)e^{-\beta E}$.
This selects a canonically most probable energy. In the thermalization regimes specified below, the thermodynamic bounds establish the corresponding macroscopic ensemble equivalence for our observables.
We denote by $\calH_L^{\bc}(\beta)$ the eigenspace of $H_L^{\bc}$ with energy $E_L^{\bc}(\beta)$, and by $\bkt{\cdot}_{L,\beta}^{\mc,\bc}$ the corresponding microcanonical expectation, namely, the normalized trace over this eigenspace.
No energy width is needed because of the large degeneracy.

Let us specify the set of macroscopic observables.
Choose a positive integer-valued function $g(L)=o(L^{1/2})$ and put $C_{g(L)}=\{0,\ldots,g(L)-1\}^d$.
For every nonempty $\Gamma\subset C_{g(L)}$, let
\eq
O_{\Gamma,L}^{\bc}=\sum_u\prod_{v\in\Gamma}Z_{u+v},\qquad
F_{\Gamma,L}^{\bc}=\frac{O_{\Gamma,L}^{\bc}}N,\lb{observables}
\en
where $u$ in the first sum runs over all of $\Lambda_L$ for $\bc=\pp$, and over those $u\in\Zb^d$ such that $u+\Gamma\subset\Lambda_L$ for $\bc=+$.
Let $\calO_L^{\bc}$ consist of these $O_{\Gamma,L}^{\bc}$.
Every elementary product has norm one, independently of its order.
For the thermalization theorems, the target equilibrium density $a_\Gamma^{\bc}(\beta)$ is the expectation of $\prod_{v\in\Gamma}Z_v$ in the unique infinite-volume Gibbs state for $\beta\le\betac(d)$, or in the plus state for $\beta>\betac(d)$.
For fixed precision $0<\delta<1$, define
\eq
P_{\neqs}^{\bc}=1-\prod_{\substack{\Gamma\subset C_{g(L)}\\ \Gamma\ne\varnothing}}
\Proj\bigl[|F_{\Gamma,L}^{\bc}-a_\Gamma^{\bc}(\beta)|\le\delta\bigr].\lb{Pneq}
\en
All factors commute. Consequently $\bra\Phi P_{\neqs}^{\bc}\ket\Phi$ is the probability that a simultaneous measurement in a normalized state $\ket\Phi$ fails at least one thermal condition. Smallness of this probability is our criterion for macroscopic thermal equilibrium (MATE).
One may also consider non-translation-invariant macroscopic observables; we restrict attention to translation sums for simplicity.

\para{Perturbation and the thermalization theorems}
The Ising Hamiltonians \rlb{Hp}, \rlb{Hplus} leave every $Z$-observable invariant. To obtain nontrivial dynamics for these observables, we add a quantum perturbation.
Following \cite{Roos}, draw $V_L=V_L^\dagger$ from normalized Lebesgue measure on $\{V:\norm V\le1\}$ in the full operator space on $\calH_{\rm tot}$, and set
\eq
\ttH_L^{\bc}=H_L^{\bc}+\lambda_L V_L,\qquad 0<\lambda_L<1.\lb{perturbation}
\en
Let $\ttcalH_L^{\bc}(\beta)$ be the spectral subspace of $\ttH_L^{\bc}$ for $(E_L^{\bc}(\beta)-2,E_L^{\bc}(\beta)+2)$.
Distinct unperturbed levels are separated by at least four, so this is precisely the cluster originating from the eigenspace $\calH_L^{\bc}(\beta)$; see Sec.~S8 of the Supplemental Material (SM) \cite{SM}.
The observables and thermal targets are not perturbed.
For every fixed $\lambda_L>0$, the perturbed energies and their nonzero gaps are nondegenerate with probability one; see Lemma~2 of \cite{Roos}.

We now state the main thermalization theorems, using $\bc=\pp$ in Theorem~\ref{THhigh} and $\bc=+$ in Theorem~\ref{THlow}.

\para{Theorem \ctl{THhigh} (thermalization in the disordered phase)}
Let $d\ge1$ and $0<\beta\le\betac(d)$, with $\beta<\infty$.
There is $c>0$ such that, for all sufficiently large $L$, one can choose an arbitrarily small deterministic $\lambda_L>0$ for which, with probability at least $1-\epsL$, where
\eq
\epsL=e^{-cR_L^{\pp}},\qquad R_L^{\pp}=\frac{L^d}{g(L)^d},\lb{ratehigh}
\en
both of the following hold. Every normalized energy eigenstate $\ket\Psi\in\ttcalH_L^{\pp}(\beta)$ satisfies $\bra\Psi P_{\neqs}^{\pp}\ket\Psi\le\epsL^2/2$.
For every normalized $\ket{\Phi(0)}\in\ttcalH_L^{\pp}(\beta)$, there are a sufficiently large $T$ and a measurable $\calA\subset[0,T]$ such that
\eq
\frac{|\calA|}{T}\le\epsL,\qquad
\bra{\Phi(t)}P_{\neqs}^{\pp}\ket{\Phi(t)}\le\epsL
\quad(t\in[0,T]\setminus\calA),\lb{dynamics}
\en
where $\ket{\Phi(t)}=e^{-i\ttH_L^{\pp}t}\ket{\Phi(0)}$.

\para{Theorem \ctl{THlow} (thermalization in the ordered phase)}
Let $d\ge2$ and $\betac(d)<\beta<\infty$.
The conclusions of Theorem~\ref{THhigh} hold with $\pp$ replaced by $+$ and
\eq
\epsL=e^{-cR_L^+},\qquad
R_L^+=\min\left\{L^{d-1},\frac{L^d}{g(L)^d}\right\}.\lb{ratelow}
\en

Here $R_L^{\pp}/L^{d/2}\to\infty$, whereas $R_L^+\ge L^{d/2}$ for all sufficiently large $L$. In $d=2$, $R_L^+=L$ for all sufficiently large $L$.
The constants $c$ depend on $d,\beta,\delta$, not on the initial state. The lower bound $L_0$ on the size $L$ also depends on the chosen function $g(L)$.

For any fixed perturbation in the stated event, a simultaneous measurement of all the observables in $\calO_L^{\bc}$ yields thermal outcomes at typical sufficiently late times, even if the initial state is far from equilibrium.
In particular, in Theorem~\ref{THlow}, a measurement of $M_L$ recovers the nontrivial spontaneous magnetization $\ms(\beta)$ within $\delta$ with probability at least $1-\epsL$.
The family includes products of every order up to $g(L)^d$, so their order may increase with $L$. If the maximal product order is fixed independently of $L$, the support restriction can instead be relaxed to $g(L)=o(L)$; see Sec.~S7 of SM \cite{SM}.

The above theorems establish full macroscopic thermalization for this class: measurement results, rather than merely expectation values, in $\ket{\Phi(t)}=e^{-it\ttH_L^{\bc}}\ket{\Phi(0)}$ agree with thermal equilibrium values within the specified precision.
The same joint measurement event controls every real linear combination $A_L=\sum_\Gamma c_{\Gamma,L}O_{\Gamma,L}^{\bc}$, with density precision $\delta\sum_\Gamma|c_{\Gamma,L}|$ and unchanged exceptional probabilities.
Thus combinations with a uniformly bounded coefficient sum are covered at any fixed precision; see Sec.~S5.1 of SM \cite{SM}.

\para{Schr\"odinger's cats}
Now take periodic boundaries, $d\ge2$, $\beta>\betac(d)$, and $0<\Delta<\ms(\beta)$.
Define the projections onto the two magnetization sectors by
\eq
\begin{gathered}
P_\pm=\Proj[|M_L\mp\ms(\beta)|\le\Delta],\\
P_{\out}=1-P_+-P_-,\qquad B_L=P_+-P_-.
\end{gathered}\lb{catproj}
\en
A normalized \emph{pure} state $\ket\Phi$ is an $\varepsilon$-balanced cat if
\eq
q:=\bra\Phi P_{\out}\ket\Phi\le\varepsilon,\qquad
|b|:=|\bra\Phi B_L\ket\Phi|\le\varepsilon.\lb{catdef}
\en
The weights of the two sectors are $p_\pm=\bra\Phi P_\pm\ket\Phi=(1-q\pm b)/2=1/2+O(\varepsilon)$, and
\eq
\ket\Phi=\sqrt{p_+}\ket{\Phi_+}+\sqrt{p_-}\ket{\Phi_-}
+\ket{\Phi_{\out}},\qquad \norm{\Phi_{\out}}^2=q,\lb{catdecomp}
\en
with normalized $\ket{\Phi_\pm}\in\operatorname{Ran}P_\pm$.
Purity makes this a coherent superposition, not a classical mixture.

\para{Theorem \ctl{THcat} (formation of balanced cats)}
For the above $d,\beta,\Delta$, there is $c>0$ such that, for all sufficiently large $L$, a deterministic $\lambda_L>0$ can be chosen with the following property.
With probability at least $1-\epsL$, where $\epsL=e^{-cL^{d-1}}$, every normalized energy eigenstate in $\ttcalH_L^{\pp}(\beta)$ is an $(\epsL^2/4)$-balanced cat.
Moreover, for every normalized pure initial state $\ket{\Phi(0)}\in\ttcalH_L^{\pp}(\beta)$, there are a sufficiently large $T$ and a measurable $\calA\subset[0,T]$ with $|\calA|/T\le\epsL$ such that $e^{-i\ttH_L^{\pp}t}\ket{\Phi(0)}$ is an $\epsL$-balanced cat for all $t\in[0,T]\setminus\calA$.

Connections between ETH, phase coexistence, and cat eigenstates have been studied in \cite{ZKH,Fratus,Serbyn}. In particular, \cite{Serbyn} examines interbranch superpositions near thermal first-order transitions in all-to-all models.
Theorem~\ref{THcat} gives a rigorous example with symmetry-related Ising phases and, additionally, a dynamical conclusion uniform over \emph{all} pure initial states in the shell. The essential new conclusion is not merely that some eigenstates are cats, but that no pure initial state in the shell is exempt from typical late-time cat formation.
No symmetry of the perturbation or of the initial state is imposed.
The coexistence of the two phases and preservation of purity make the conclusion necessary once the two thermal weights are recovered: the pure state cannot become an incoherent mixture. Even an initially phase-localized pure state must develop the almost balanced superposition.

A magnetization measurement gives a value near $+\ms$ or $-\ms$ with almost equal probabilities, not a value near zero.
Each branch also satisfies the appropriate plus- or minus-phase thermal conditions for the growing family in \rlb{observables}, with error scale \rlb{ratelow}, as proved in Sec.~S6 of SM \cite{SM}.
Thus the conditional state associated with either phase outcome of a coarse-grained magnetization measurement is in the corresponding macroscopic thermal equilibrium.
Such macroscopic superpositions can also be fragile under suitable weak environmental couplings \cite{ShimizuMiyadera}; the resulting open-system dynamics is not addressed here.

A symmetry-breaking quench can convert macroscopic magnetization fluctuations into an extensive energy spread, obstructing thermalization to a single narrow-energy ensemble \cite{ReimannQuench}.
This differs from our setting, where the initial state belongs to an energy shell of the Hamiltonian generating the dynamics.

\para{Thermodynamic bounds}
Essential ingredients of the theorems, apart from the theory of Roos, Sugimoto, Teufel, Tumulka, and Vogel \cite{Roos}, are the following bounds for the equilibrium states of the unperturbed model, which we call the thermodynamic bounds \cite{Tasaki2016}.

\prop{Proposition \ctl{PRthermal}}
Under the assumptions of Theorems~\ref{THhigh} and \ref{THlow}, respectively, there is $c_0>0$ such that
\eq
\bkt{P_{\neqs}^{\bc}}_{L,\beta}^{\mc,\bc}\le e^{-c_0R_L^{\bc}}.\lb{thermobound}
\en
\prop{Proposition \ctl{PRcat}}
Under the assumptions of Theorem~\ref{THcat}, there is $c_0>0$ such that
\eq
\bkt{P_{\out}}_{L,\beta}^{\mc,\pp}\le e^{-c_0L^{d-1}},\qquad
\bkt{B_L}_{L,\beta}^{\mc,\pp}=0.\lb{catbound}
\en

Equation~\rlb{thermobound} implies that a uniformly drawn pure state in the energy shell is in MATE with overwhelming probability. The corresponding typicality persists in the slightly perturbed shell, since its projection approaches the unperturbed one in norm; see Sec.~S5 of SM \cite{SM}.

The magnetization estimates underlying these propositions are highly nontrivial results of rigorous statistical mechanics \cite{Pfister,Ioffe,Pisztora,BodSlab,BodStates,ADS,DST}; the precise inputs are given in Sec.~S1 of SM \cite{SM}.
The ordered-phase scale is a surface scale: a macroscopic minority-phase droplet costs interfacial, not bulk, free energy.
Two additional arguments extend the bounds to all the products in \rlb{observables}. For bounded-order products, monotone coupling to independent plus-boundary blocks reduces concentration to magnetization concentration. For high-order products, conditioning on a color class makes individual spins independent; a concentration inequality for overlapping functions of independent variables \cite{GLSS} gives an exponent of order $N/|\Gamma|$.
Summing over at most $2^{g(L)^d}$ products gives the restriction $g(L)^{2d}=o(N)$.
At criticality, the first argument uses $\ms(\betac)=0$ and fixed-precision magnetization large deviations; it does not assume Gaussian critical fluctuations.
The product estimates and the simultaneous bound are derived in Secs.~S2--S5 of SM \cite{SM}.

The bounds discussed in the preceding paragraph are stated for the canonical ensemble.
The passage to the microcanonical ensemble is elementary.
Let $r_L^{\bc}=O(N)$ be the number of distinct eigenvalues of $H_L^{\bc}$.
Noting that $Z_L^{\bc}(\beta)=\Tr[e^{-\beta H_L^{\bc}}]\le r_L^{\bc}D_L^{\bc}(E_L^{\bc}(\beta))e^{-\beta E_L^{\bc}(\beta)}$, one readily finds $\bkt{Q}_{L,\beta}^{\mc,\bc}\le r_L^{\bc}\bkt{Q}_{L,\beta}^{\can,\bc}$ for any $Q\ge0$ \cite{Tasaki2016}.
The polynomial factor $r_L^{\bc}$ is absorbed by all the stated exponential bounds; see Sec.~S5 of SM \cite{SM} for the full argument.

\para{Ideas of the proof}
Enumerate the classical configurations in a target energy eigenspace of dimension $D$ as $\bssigma_1,\ldots,\bssigma_D$. The Fourier basis
\eq
\ket{\psi_\ell}=\frac1{\sqrt D}\sum_{j=1}^D e^{2\pi i j\ell/D}\ket{\bssigma_j}
\quad(\ell=1,\ldots,D)\lb{Fourier}
\en
has exactly the microcanonical expectation for every diagonal observable.
Thus Proposition~\ref{PRthermal} supplies the basis required by \cite{Roos}, yielding Theorems~\ref{THhigh} and \ref{THlow}.
For cats, spin flip gives the zero microcanonical mean in \rlb{catbound}. Haar concentration within the exponentially large energy eigenspace controls all diagonal \emph{and off-diagonal} matrix elements of $B_L$.
Nondegenerate energy gaps then bound the long-time mean of $|\bra{\Phi(t)}B_L\ket{\Phi(t)}|^2$ for every initial state.
This extra step is indispensable: concentration in the union of the two phases alone does not establish their balanced weights.
The quantum arguments are given in the End Matter.

\para{Discussion}
We have proved thermalization and the formation of Schr\"odinger's cats under unitary time evolution in the Ising model with a weak random perturbation. The results provide, to our knowledge, the first rigorous example of macroscopic thermalization from every initial state in an energy shell of an interacting lattice model, with equilibrium characterized by the measurement results of an increasing family of macroscopic observables. Unlike the spatial-density equilibration in the free-fermion examples, the measured correlations recover their temperature-dependent thermal values, including spontaneous magnetization. The general treatment of degeneracy and weak random perturbations is due to Roos, Sugimoto, Teufel, Tumulka, and Vogel \cite{Roos}; the present application establishes these thermal conclusions throughout the disordered and ordered phases, as well as balanced-cat formation from every pure initial state in a shell. We believe this provides a concrete step towards a quantum-mechanical foundation of equilibrium statistical mechanics.

We must admit, however, that the full-space random perturbation $V_L$ is unrealistic. Almost surely, it has nonzero matrix elements between every pair of classical basis states in $\calH_L^{\bc}(\beta)$, and its compression to this shell has a Haar-distributed eigenbasis. The proof exploits this mixing within the highly degenerate shell, not a local transport mechanism.
No useful lower bound on the admissible perturbation strength is obtained. Relaxation-time estimates may be sought using the random-basis methods of \cite{GHTfast,ReimannTime}, but we do not pursue their adaptation to the perturbation-induced spectrum: the resulting scale would have limited direct physical significance.
It remains nontrivial, and pedagogically important, that unitary evolution alone suffices to thermalize every initial state in the shell of this interacting system. The perturbation is drawn once; no stochastic evolution or averaging over initial states is involved. See Section~2.4 of \cite{Roos} for further discussion of generic perturbations.

A central challenge is whether a realistic short-range perturbation yields similar ETH and thermalization results. In $d\ge2$ with periodic boundaries, adding a nonzero transverse field $\sum_u X_u$ \cite{Chiba}, or a nearest-neighbor exchange $\sum_{\langle u,v\rangle}X_uX_v$ \cite{STcharges}, removes nontrivial local conserved quantities in the senses established there. These interactions are extensive, unlike our norm-one perturbations; absence of local conservation laws alone does not imply ETH or thermalization. Preliminary rigorous results on operator growth and complex-time evolution in related local models appear in Appendices A.3 and A.4 of \cite{STcharges}.


\medskip
\begin{samepage}
\begin{acknowledgments}
I thank Stefan Teufel and Roderich Tumulka for valuable discussions, Aernout C.~D.~van Enter for a comment on the scale of the perturbation, Akira Shimizu for ongoing discussions on thermal equilibrium, and Maksym Serbyn for enlightening discussions on the relation between first-order phase transitions and Schr\"odinger-cat states.
This work is supported by JSPS Grants-in-Aid for Scientific Research Nos.~22K03474 and 25K07171.
\end{acknowledgments}
\end{samepage}

\para{Use of generative artificial intelligence}
I prepared the original 2024 note \cite{Earlier}, where the essential setup and idea are presented, without AI assistance. The present substantially expanded work used ChatGPT Pro (versions 5.6 and 6) to identify an incorrect use of a large-deviation result in \cite{Earlier}, locate references, develop and check proofs, and prepare the exposition. The setting and principal conclusions were proposed by me, except for the applicability of the theorems at the critical temperature and the thermal characterization of the cat branches, which were suggested by ChatGPT Pro 6 before I asked about them. In particular, the proof of Lemma~S1, which derives surface-order concentration for bounded-sensitivity observables from magnetization large deviations by a monotone block-coupling argument, was developed by ChatGPT. I have checked the final manuscript and take full responsibility for all arguments, references, and text.

\para{Data availability}
No numerical or experimental data were generated or analyzed in this work. The arguments are contained in the article and its Supplemental Material.

\par\bigskip
\section*{End Matter}
\label{sec:endmatter}
We first use the theory in \cite{Roos} to turn the thermodynamic bound into eigenstate thermalization and then thermalization for every initial state.
For cat formation, we additionally control the relative weights of the two magnetization sectors; this requires off-diagonal matrix elements as well as their diagonal counterparts.
\para{From the thermodynamic bound to eigenstate thermalization}
Fix $L,\beta,\bc$, abbreviate the target shell projection by $P_0$, its dimension by $D$, and set $Q=P_{\neqs}^{\bc}$.
By Proposition~\ref{PRthermal}, after choosing a smaller $c$ in the theorems,
\eq
q_0:=\frac{\Tr(P_0Q)}D\le e^{-12cR},\qquad
\epsL=e^{-cR},\qquad R=R_L^{\bc}.\lb{EMq}
\en
The Fourier basis \rlb{Fourier} consists of eigenvectors of the unperturbed Hamiltonian and has $\bra{\psi_\ell}Q\ket{\psi_\ell}=q_0$.
Apply Proposition~4 of \cite{Roos} to this single eigenspace with equilibrium error $q_0$, bounded as in \rlb{EMq}.
If $(\ket{\varphi_a})_{a=1}^D$ is a Haar-random orthonormal basis and $\eta=\epsL^2/32$, that proposition gives
\eq
\begin{split}
&\Prob\{\exists a:\bra{\varphi_a}Q\ket{\varphi_a}>q_0+\eta\}\\
&\hspace{10mm}\le 2D\exp[-\CR\eta^3 e^{12cR}],\qquad
\CR=\frac2{9\pi^3}.\lb{EMbasis}
\end{split}
\en
Its hypothesis $\CR\eta^3\ge e^{-12cR}$ holds for large $L$.
Since $D\le2^N$ and $R/\log N\to\infty$, the right-hand side is smaller than $\epsL/4$ for large $L$.
When $Q$ vanishes on the shell the same conclusion is immediate.

The compressed random operator $P_0V_LP_0$ has an absolutely continuous distribution invariant under all unitaries on $\operatorname{Ran}P_0$.
Its spectrum is simple almost surely, and its eigenbasis, up to phases and ordering, is Haar distributed.
For each such $V_L$, degenerate perturbation theory implies that the eigenvectors of $H_L^{\bc}+\lambda V_L$ in the isolated cluster converge, as $\lambda\downarrow0$, to this basis.
Thus, for almost every $V_L$, these finitely many changes stay below $\eta$ throughout a sufficiently small positive interval of $\lambda$.
Choose a deterministic $\lambda_L$ so that this holds for every $0<\lambda\le\lambda_L$ outside a set of probability at most $\epsL/4$.
This is the perturbative step in the proof of Theorem~1 of \cite{Roos}; it makes no claim of a size-uniform perturbative radius.
For large $L$, $q_0+2\eta\le\epsL^2/2$, proving the eigenstate statements in Theorems~\ref{THhigh} and \ref{THlow}.

\para{Thermalization for every initial state}
Let $\Pi_a$ be the distinct-energy spectral projections of $\ttH_L^{\bc}$ within its shell.
The eigenstate statement means $\norm{\Pi_aQ\Pi_a}\le\epsL^2/2$, even if a level is degenerate.
For any normalized $\ket{\Phi(0)}$ in the perturbed shell,
\eq
\begin{split}
\lim_{T\uparrow\infty}\frac1T\int_0^T\bra{\Phi(t)}Q\ket{\Phi(t)}\,dt
&=\sum_a\bra{\Phi(0)}\Pi_aQ\Pi_a\ket{\Phi(0)}\\
&\le\epsL^2/2.\lb{EMtime}
\end{split}
\en
For every sufficiently large $T$ the integral average is at most $\epsL^2$.
Taking $\calA=\{t\in[0,T]:\bra{\Phi(t)}Q\ket{\Phi(t)}>\epsL\}$ and applying Markov's inequality proves \rlb{dynamics}.
The good event for $V_L$ is independent of the initial state; $T$ and $\calA$ may depend on it.
This is the dynamical argument of Proposition~1 of \cite{Roos}.

\para{Exponential shell dimension and spin flip}
At fixed finite temperature, the shell contains exponentially many states; this will make the concentration estimate in the cat proof effective.
For either boundary condition, choose a set $I_L$ with no two nearest neighbors and $|I_L|\ge N/(2d+1)$. A greedy construction retains one site at each step and removes it and its at most $2d$ neighbors from further consideration.
Starting from the all-plus ground configuration, flipping any subset of $I_L$ costs exactly $4d$ per spin, additively, since no bond joins two flipped sites.
Write $D=\dim\calH_L^{\bc}(\beta)$ and let $E_g$ be the ground energy. Then
\eq
\begin{split}
Z_L^{\bc}(\beta)&\ge e^{-\beta E_g}(1+e^{-4d\beta})^{|I_L|},\\
Z_L^{\bc}(\beta)&\le r_L^{\bc} D e^{-\beta E_L^{\bc}(\beta)}.
\end{split}\lb{EMdim0}
\en
This gives $D\ge e^{s_\beta N}$ for all large $L$, since $E_L^{\bc}(\beta)\ge E_g$ and $r_L^{\bc}=O(N)$. For example, one may take $s_\beta=\log(1+e^{-4d\beta})/[2(2d+1)]$.
This is where fixed \emph{finite} temperature matters; the ground eigenspace on the torus has dimension two and is not covered.
For periodic boundaries, global spin flip $U_L=\prod_{u\in\Lambda_L}X_u$ preserves each periodic exact-energy eigenspace and interchanges $P_+$ and $P_-$.
Hence $\Tr(P_0B_L)=0$, without any symmetry assumption on $V_L$.

\para{Proof of Theorem~\ref{THcat}}
Use Proposition~\ref{PRcat} with a smaller $c$ so that $q_0=\Tr(P_0P_{\out})/D\le\epsL^{12}$, where $\epsL=e^{-cL^{d-1}}$.
Equation~(40) of \cite{Roos}, in the proof of its Proposition~2, gives, for $\eta>0$ and any self-adjoint $A$ on a $D$-dimensional space that is not a multiple of the identity,
\eq
\Prob\left\{\left|\bra\varphi A\ket\varphi-\frac{\Tr A}{D}\right|>\eta\right\}
\le2\exp\left[-\frac{\CR\eta^2D}{\Delta_A^2}\right],\lb{EMHaar}
\en
where $\ket\varphi$ is Haar distributed and $\Delta_A>0$ is the spectral width.
If $A$ is a constant multiple of the identity, the probability on the left is zero for every $\eta>0$.
Apply this to the compressions of $P_{\out}$ and $B_L$, whose spectral widths are at most one and two, respectively.
For a Haar basis and $a\ne b$, each vector $(\ket{\varphi_a}\pm\ket{\varphi_b})/\sqrt2$ and $(\ket{\varphi_a}\pm i\ket{\varphi_b})/\sqrt2$ is also Haar distributed.
Polarization and a union bound therefore bound all off-diagonal entries of $B_L$ as well as its diagonal entries.
With $\eta=\epsL^2/64$, the exceptional probability is at most
\eq
2D e^{-\CR\eta^2D}+8D^2e^{-\CR\eta^2D/4}=o(\epsL),\lb{EMcatfail}
\en
using $D\ge e^{s_\beta N}$ and $d\ge2$.
Repeating the perturbative passage above, choose $\lambda_L$ so that, with probability at least $1-\epsL$,
\eq
\bra{\Psi_a}P_{\out}\ket{\Psi_a}\le\epsL^2/4,\qquad
|\bra{\Psi_a}B_L\ket{\Psi_b}|\le\epsL^2/4\lb{EMentries}
\en
for every pair of perturbed shell eigenstates.
We also use the almost-sure nondegeneracy of energies and their nonzero gaps noted in the Letter.

Write $\ket{\Phi(0)}=\sum_a c_a\ket{\Psi_a}$ and set $q(t)=\bra{\Phi(t)}P_{\out}\ket{\Phi(t)}$ and $b(t)=\bra{\Phi(t)}B_L\ket{\Phi(t)}$.
Denote infinite-time averages by an overline and write $(B_L)_{ab}=\bra{\Psi_a}B_L\ket{\Psi_b}$.
Then \rlb{EMentries} implies
\eqa
\overline{q(t)}&\le\epsL^2/4,\lb{EMqtime}\\
\overline{|b(t)|^2}
&=\left|\sum_a |c_a|^2 (B_L)_{aa}\right|^2
+\sum_{a\ne b}|c_a|^2|c_b|^2|(B_L)_{ab}|^2\nl
&\le\epsL^4/8.\lb{EMbtime}
\ena
The second identity uses nondegenerate nonzero gaps.
Choose sufficiently large $T$ so that the finite averages are at most $\epsL^2/2$ and $\epsL^4/4$.
The union of the sets $q(t)>\epsL$ and $|b(t)|>\epsL$ then occupies a fraction at most $\epsL/2+\epsL^2/4\le\epsL$ of $[0,T]$.
This proves the dynamical statement for every initial pure state.
The coherence between the two sectors is also explicit:
$\norm{P_+\ket{\Phi(t)}\bra{\Phi(t)}P_-}_1=\sqrt{p_+p_-}=1/2-O(\epsL)$ at these times, where $p_\pm=\bra{\Phi(t)}P_\pm\ket{\Phi(t)}$.
\qedm

\clearpage
\onecolumngrid
\setcounter{page}{1}
\renewcommand{\thepage}{S\arabic{page}}
\setcounter{secnumdepth}{2}
\setcounter{section}{0}
\renewcommand{\thesection}{S\arabic{section}}
\setcounter{subsection}{0}
\renewcommand{\thesubsection}{\thesection.\arabic{subsection}}
\makeatletter
\renewcommand{\p@subsection}{}
\makeatother
\setcounter{equation}{0}
\renewcommand{\theequation}{S\arabic{equation}}
\renewcommand{\theHequation}{supp.\arabic{equation}}
\renewcommand{\theHsection}{supp.\arabic{section}}
\renewcommand{\theHsubsection}{supp.\arabic{section}.\arabic{subsection}}
\begin{center}
{\large\bf Supplemental Material}\\[5pt]
{\bf Full macroscopic thermalization and the formation of Schr\"odinger's cats\\
by unitary time evolution in the weakly perturbed Ising model}\\[4pt]
Hal Tasaki
\end{center}
\noindent
We prove the classical estimates used in Propositions~\ref{PRthermal} and \ref{PRcat}, and the thermal characterization of the cat branches.
The quantum transfer is proved in the End Matter.
All statements use $H=-\sum ZZ$ with the boundary terms in \rlb{Hplus}, standard norm-one Pauli products, fixed finite $\beta$, and fixed positive measurement precision.
Constants may depend on $d,\beta$ and these precisions, and may change between displays.
No estimate uniform as $\beta\uparrow\infty$ or as the precision tends to zero is asserted.
The reference numbers are those of the Letter.

The Supplemental Material is organized as follows. Section~\ref{sec:input} supplies the magnetization estimates on which the proofs rest: volume-order concentration when the spontaneous magnetization vanishes, including criticality, and surface-order concentration in the ordered phase. It identifies the precise rigorous Ising results used and explains the passage from boundary-uniform random-cluster estimates to the periodic two-peak bound.

Sections~\ref{sec:comparison}--\ref{sec:highorder} extend these inputs from magnetization to spin products. Section~\ref{sec:comparison} proves a comparison lemma for arbitrary functions whose change under one spin flip is bounded by a fixed constant divided by $N$, using a monotone coupling to independent plus-boundary blocks. Section~\ref{sec:targets} then replaces finite-volume means by the infinite-volume Gibbs targets, uniformly for products of bounded order even when the distances between their sites grow. Section~\ref{sec:highorder} handles the complementary, high-order products by conditioning on one sublattice and applying the read-$k$ concentration inequality. This last step preserves norm-one Pauli normalization and requires neither uniqueness nor decay of correlations.

Section~\ref{sec:conditioning} combines the two product estimates and controls the number of observables, which yields the support condition $g(L)=o(\sqrt L)$. The elementary comparison between the canonical measure and the microcanonical ensemble at the selected exact energy then proves Propositions~\ref{PRthermal} and~\ref{PRcat}. The same section derives typicality in the unperturbed and perturbed energy shells, and its final subsection proves the measurement statement for linear combinations.

Section~\ref{sec:branches} proves that the two branches of a cat satisfy their respective plus- and minus-phase thermal conditions. It first establishes the required phase-resolved thermodynamic bound, then applies the quantum argument in the End Matter, and finally formulates the conclusion for the normalized branches selected by a magnetization measurement. Section~\ref{sec:longer} defines the restricted observable family for a fixed maximal product order and $g(L)=o(L)$, states its thermalization and thermal-branch theorems explicitly, and proves them with volume- or surface-order errors. Section~\ref{sec:arithmetic} verifies the energy-level spacing, the number of distinct energies, and the properties and limitations of the perturbation ensemble.

Thus Sections~\ref{sec:input}--\ref{sec:conditioning}, together with the energy arithmetic in Sec.~\ref{sec:arithmetic}, provide the equilibrium input for the three theorems of the Letter; Secs.~\ref{sec:branches} and~\ref{sec:longer} establish the two extensions mentioned there. The passage from these inputs to unitary dynamics is contained in the End Matter, rather than repeated here.

\section{Classical equilibrium input and the periodic two-peak bound}
\label{sec:input}
We collect the finite-volume magnetization bounds needed for both thermalization and cat formation. The main point requiring a separate explanation is the periodic two-peak estimate in the ordered phase, especially in dimensions $d\ge3$.

Let $\mu_{L,\beta}^{\bc}$ denote the classical Gibbs measure of $H_L^{\bc}$, and write $\mu_\beta^+$ for the infinite-volume plus state. Its one-site mean is $\ms=\ms(\beta)$.
We identify diagonal operators with functions of $\bssigma\in\{-1,1\}^{\Lambda_L}$, and write $M(\bssigma)=N^{-1}\sum_u\sigma_u$.
The elementary monotonicity and infinite-volume statements used below are discussed in \cite{FV}.

\subsection{Fixed-precision volume bounds when the spontaneous magnetization vanishes}
We derive a volume-order bound for any fixed nonzero magnetization deviation from differentiability of the Massieu potential density (often called the pressure in the mathematical literature) at zero field. The argument includes the critical point without requiring a quantitative critical fluctuation estimate.
The Massieu potential density with a dimensionless field is
\eq
p_\beta(h)=\lim_{L\uparrow\infty}\frac1N\log\sum_{\bssigma}
\exp\left[-\beta H_L^{\bc}(\bssigma)+h\sum_u\sigma_u\right].\lb{massieu}
\en
The limit is independent of $\bc=\pp,+$, since changing boundary bonds changes the logarithm of the partition function by $O(L^{d-1})$.
The right derivative at zero is $p_\beta'(0+)=\ms$.
When $\beta\le\betac(d)$, including criticality, $\ms=0$; the critical continuity results are in \cite{ADS,DST}.
For $d=1$, $\ms=0$ at every finite $\beta$.
For any fixed $a>0$, choose $h>0$ so small that $p_\beta(h)-p_\beta(0)<ha/2$.
Exponential Markov bounds and the convergence in \rlb{massieu} give, for either boundary condition and all sufficiently large $L$,
\eq
\mu_{L,\beta}^{\bc}(M>a)
\le e^{-haN}\frac{Z_{L,\beta}^{\bc}(h)}{Z_{L,\beta}^{\bc}(0)}
\le e^{-haN/4}.\lb{massieutail}
\en
The negative tail follows by using $-h$. Thus
\eq
\mu_{L,\beta}^{\bc}(|M|>a)\le C_a e^{-c_aN}
\quad(\beta\le\betac(d)).\lb{maghigh}
\en
This is a fixed-precision estimate. No Gaussian rate uniform in $a$, or claim about critical fluctuation scaling, is involved.
The same argument with $p_\beta'(0+)=\ms>0$ bounds $M>\ms+a$ by a volume-order exponential in the ordered phase.

\subsection{Surface-order estimates in the ordered phase}
We now use the surface-order Ising estimates to control deviations from the spontaneous magnetization. For periodic boundaries the target is the pair of values $\pm\ms$, rather than a single phase value.
For early work on surface-order large deviations and phase separation, see \cite{Pfister}; the later estimates used here cover the whole ordered phase.
For $d\ge2$ and every fixed $\betac(d)<\beta<\infty$, the required estimates are
\eqa
\mu_{L,\beta}^+(|M-\ms|>a)&\le C_a e^{-c_aL^{d-1}},\lb{magplus}\\
\mu_{L,\beta}^{\pp}(\,||M|-\ms|>a)&\le C_a e^{-c_aL^{d-1}}\qquad(a>0).\lb{magperiodic}
\ena
In $d=2$, the nontrivial lower-tail bound in \rlb{magplus} is Theorem~1.1 of Ioffe \cite{Ioffe}; the remark immediately following that theorem states the extension to periodic boundary conditions used for \rlb{magperiodic}.
The upper tails beyond $\ms+a$ have the stronger volume-order bound just discussed.

For $d\ge3$, the plus-boundary result follows from Theorem~1.1 of Pisztora \cite{Pisztora}, after removing both qualifications in its original formulation: the slab threshold agrees with the Ising critical threshold \cite{BodSlab}, and the free/wired FK limits and the relevant magnetization limits agree away from criticality \cite{BodStates}.
Section~2.3 of \cite{BodStates} explicitly explains why the two hypotheses of the coarse-graining theory then hold throughout $\beta>\betac$.
Both results are needed here. Differences between fixing spins on the boundary of a box and fixing its exterior neighbors are removed by adding one layer of fixed spins; the change in magnetization density is $O(L^{-1})$ and is absorbed in the fixed precision.
Pisztora uses an Ising interaction equal to $-\sigma_u\sigma_v/2$ up to an additive constant; his inverse temperature is therefore $2\beta$ in our convention.
The corresponding FK bond parameter is $p=1-e^{-2\beta}$.

For completeness, we explain the periodic passage from the FK estimate, rather than identifying it with the literal plus-boundary spin statement.
Theorem~1.2 of \cite{Pisztora} is uniform over arbitrary partitions wiring boundary vertices of a cube together.
For every sufficiently small fixed $a>0$, it gives a fixed $\ell<\infty$ such that, except with probability $Ce^{-cL^{d-1}}$, the internal open bonds have a unique largest cluster $C_*$, this cluster crosses the cube, and
\eq
\bigl||C_*|/N-\ms\bigr|<a,\qquad
\sum_{\substack{C\ne C_*\\ \operatorname{diam}C>\ell}}|C|<aN.\lb{FKgood}
\en
The equality of the limiting free and wired cluster densities, both equal to $\ms$, is used in the first inequality.

Cut the torus into an ordinary cube and condition on the wrapping bonds.
Their open connections induce precisely a boundary wiring partition of the kind allowed by that theorem, so \rlb{FKgood} holds uniformly under the conditioning.
Small internal clusters touching the boundary occupy at most $C_d\ell L^{d-1}$ sites: every such cluster lies in a layer of thickness $\ell$.
Discard these sites and those in the intermediate clusters in \rlb{FKgood}.
Outside $C_*$, at most $aN+O(\ell L^{d-1})$ discarded sites can join its torus extension through the wrapping bonds.
All other small clusters are internal, remain separate torus clusters, and have size at most $C_d\ell^d$.
Conditional on the full FK configuration, their Ising colors are independent fair signs.
The sum of their squared sizes is at most $C_d\ell^d N$, and hence their total spin exceeds $aN$ in absolute value with probability at most $2e^{-c_da^2N/\ell^d}$.
The large torus cluster also has a fair color.
Thus the magnetization is within $C a+o(1)$ of either $\ms$ or $-\ms$, with the asserted surface-order exceptional probability.
Choosing $a$ smaller proves \rlb{magperiodic} for every desired fixed tolerance.

\section{A comparison lemma for bounded-sensitivity functions}
\label{sec:comparison}
Our aim is to transfer magnetization concentration to any real function with uniformly small single-spin sensitivity. The comparison with independent plus-boundary blocks allows us to do this without a mixing assumption, including at criticality and throughout the ordered plus phase.

For a real function $F$ on the configurations, let
\eq
\osc_u(F)=\sup_{\bssigma}|F(\bssigma)-F(\bssigma^u)|,\lb{osc}
\en
where $\bssigma^u$ differs only at $u$.
The following deduction from the magnetization estimates will be useful.

\para{Lemma S1}
Fix $K>0$ and $\delta>0$. If
\eq
\max_u\osc_u(F_L)\le K/N,\lb{sensitivity}
\en
then, for all sufficiently large $L$,
\eq
\mu_{L,\beta}^{\bc}(|F_L-\mu_{L,\beta}^{\bc}F_L|>\delta)
\le C e^{-cA_L^{\bc}},\qquad
A_L^{\bc}=\begin{cases}
N,&\beta\le\betac(d),\quad\bc=\pp,+,\\
L^{d-1},&\beta>\betac(d),\quad\bc=+.
\end{cases}\lb{comparisonbound}
\en
The constants are uniform over all real $F_L$ satisfying \rlb{sensitivity}.

\para{Proof}
Put $a=\min\{\delta/(8K),1/8\}$.
Choose a fixed block side $\ell$ so large that the spatially averaged magnetization in a plus-boundary $\ell$-box is at most $\ms+a/2$.
Partition most of $\Lambda_L$ into such boxes and the remaining $O(\ell L^{d-1})$ sites into smaller boxes.
Let
\eq
\nu_L=\bigotimes_{C}\mu_{C,\beta}^+.\lb{blockmeasure}
\en
For sufficiently large $L$, $\nu_L M\le\ms+a$.
The product measure $\nu_L$ stochastically dominates $\mu_{L,\beta}^+$ and also $\mu_{L,\beta}^{\pp}$.
Indeed, replacing every spin outside a block by a fixed plus spin raises each conditional local field; the monotone coupling criterion gives a coupling $(\bssigma,\bseta)$ with the original measure for $\bssigma$, product measure for $\bseta$, and $\sigma_u\le\eta_u$ at every site.
The number of disagreements in this coupling is exactly $N[M(\bseta)-M(\bssigma)]/2$. Therefore
\eq
|F_L(\bseta)-F_L(\bssigma)|\le\frac K2[M(\bseta)-M(\bssigma)].\lb{Hamming}
\en
For plus boundaries $\mu_{L,\beta}^+M\ge\ms$; for periodic boundaries in the regime used here, both $\ms$ and $\mu_{L,\beta}^{\pp}M$ are zero.
Taking expectations in \rlb{Hamming} gives
\eq
|\nu_LF_L-\mu_{L,\beta}^{\bc}F_L|\le Ka/2.\lb{blockmean}
\en

The block variables under $\nu_L$ are independent. Changing one block changes $M$ by at most $2\ell^d/N$ and $F_L$ by at most $K\ell^d/N$.
The bounded-differences inequality consequently gives
\eqa
\nu_L(M>\ms+2a)&\le e^{-a^2N/(2\ell^d)},\lb{blockM}\\
\nu_L(|F_L-\nu_LF_L|>\delta/2)&\le2e^{-\delta^2N/(2K^2\ell^d)}.\lb{blockF}
\ena
The remaining event $M(\bssigma)<\ms-a$ has the bound \rlb{magplus} or \rlb{maghigh}.
Outside these three events, \rlb{Hamming} is at most $3Ka/2$.
Together with \rlb{blockmean} and the $\delta/2$ block deviation, this is smaller than $\delta$.
The union bound proves \rlb{comparisonbound}.
\qedm

\section{Uniform targets for products of bounded order}
\label{sec:targets}
The preceding lemma gives concentration about a finite-volume mean, whereas the Letter uses infinite-volume Gibbs expectations as thermal targets. We show that these means approach the specified targets uniformly for products of bounded order, even when the support diameter grows as $o(L)$.

For $\Gamma\subset C_{g(L)}$, let $\calU_{\Gamma,L}^{\bc}$ be the set of translations used in \rlb{observables}: $\calU_{\Gamma,L}^{\pp}=\Lambda_L$ and $\calU_{\Gamma,L}^+=\{u\in\Zb^d:u+\Gamma\subset\Lambda_L\}$.
Set $k=|\Gamma|$ and $n_{\Gamma,L}=|\calU_{\Gamma,L}^+|$.
The geometric bounds
\eq
(L-g(L)+1)^d\le n_{\Gamma,L}\le N,\qquad
\osc_u(F_{\Gamma,L}^{\bc})\le\frac{2k}N\lb{geom}
\en
hold uniformly in $\Gamma$.
Fix a maximal order $k_0$ and assume only $g(L)=o(L)$ in this section.

With plus boundaries, couple the finite-volume plus state above the restriction of the infinite-volume plus state.
A product changes by at most twice the number of disagreements on its support.
A given site occurs in at most $k$ of its translated supports.
Consequently
\eq
\left|\mu_{L,\beta}^+F_{\Gamma,L}^+-a_\Gamma^+(\beta)\right|
\le\frac{k}{N}\sum_{u\in\Lambda_L}
[\mu_{L,\beta}^+(\sigma_u)-\ms]+1-\frac{n_{\Gamma,L}}N=o(1),\lb{targetplus}
\en
uniformly in $k\le k_0$ and in the positions of the support sites.
Convergence of the mean magnetization and $g(L)/L\to0$ suffice; exponential mixing is not needed.

For periodic boundaries with $\beta\le\betac$, compare both the torus measure, viewed on the cut cube, and the restriction of the infinite-volume state with the plus-boundary measure on that cube.
Both measures have zero one-site means and are stochastically dominated by the plus measure.
Telescoping over disagreements and then over the two comparisons gives
\eq
\left|\mu_{L,\beta}^{\pp}F_{\Gamma,L}^{\pp}-a_\Gamma^{\pp}(\beta)\right|
\le2k\,\mu_{L,\beta}^+M
+2\left[1-\frac{(L-g(L)+1)^d}N\right]=o(1).\lb{targetp}
\en
The last term bounds the translates that wrap around the cut and therefore do not have the infinite-volume target.
This argument also holds at criticality.
Applying Lemma S1 with $K=2k_0$ and a slightly smaller precision now yields
\eq
\mu_{L,\beta}^{\bc}(|F_{\Gamma,L}^{\bc}-a_\Gamma^{\bc}(\beta)|>\delta)
\le C e^{-c A_L^{\bc}},\qquad 1\le|\Gamma|\le k_0,\lb{boundedorder}
\en
uniformly in the allowed supports.

\section{High-order norm-one products}
\label{sec:highorder}
We now control products whose order is not bounded independently of $L$. Finite-temperature randomness makes their thermal expectations small, and a conditional concentration argument also controls the fluctuations of their translation averages.

The argument uses only finite temperature and the nearest-neighbor structure, not uniqueness or decay of correlations.
For plus boundary conditions with any $L$, or periodic boundary conditions with even $L$, use the standard bipartite partition $\Lambda_L=\Lambda_L^0\cup\Lambda_L^1$, where $\Lambda_L^j=\{u\in\Lambda_L:\sum_i u_i\equiv j\pmod2\}$.
There are no interactions within either sublattice.
Conditional on the spins in one sublattice, the spins in the other are independent, and each conditional mean has absolute value at most
\eq
\rho=\tanh(2d\beta)<1.\lb{rho}
\en
This also holds at the boundary in the plus case: the fixed exterior spins contribute only one-site fields, and every dynamical site still has $2d$ neighbors in total.

For a product on $k$ sites, at least $k/2$ of them belong to one sublattice.
Conditioning on the other sublattice gives
\eq
|\mu(\sigma_\Gamma)|\le b_k:=\rho^{k/2},\qquad
\sigma_\Gamma=\prod_{v\in\Gamma}\sigma_v,\lb{smallmean}
\en
for either finite-volume geometry just described and for every infinite-volume Gibbs state, using the corresponding bipartition of $\Zb^d$.
Importantly, this is a small \emph{expectation}, not a small norm: $\norm{Z_\Gamma}=1$ for every $k$, where $Z_\Gamma=\prod_{v\in\Gamma}Z_v$.

We also need a probability estimate, since \rlb{smallmean} alone does not imply concentration.
A family of functions of independent inputs is called read-$k$ if each input occurs in at most $k$ functions. Theorem~1.1 of \cite{GLSS}, after replacing a sign by its associated indicator and using the binary relative-entropy bound, implies the following: for read-$k$ signs $W_1,\ldots,W_r$,
\eq
\Prob\left(\left|\sum_{j=1}^r[W_j-\E W_j]\right|>t\right)
\le2\exp\left[-\frac{t^2}{2kr}\right].\lb{readk}
\en
The means need not be equal.

Fix $\Gamma$ of order $k$ and abbreviate its admissible translation set to $\calU$, of size $n$.
For each translation $u$, assign the sign $W_u=\prod_{v\in\Gamma}\sigma_{u+v}$ to a sublattice containing at least $k/2$ of its support sites, breaking ties in a fixed way.
The assignment depends only on the support, not on the spins.
Let $\calU_j$ be the translations assigned to $\Lambda_L^j$, $n_j=|\calU_j|$, and
\eq
S_j=\sum_{u\in\calU_j}W_u,\qquad
J_j=\E(S_j\mid\bssigma_{\Lambda_L\setminus\Lambda_L^j}),\qquad j=0,1.\lb{sublatticesums}
\en
Conditional on the other sublattice, the signs in $S_j$ form a read-$k$ family: any one spin belongs to at most $k$ translated copies of $\Gamma$.
Their conditional means have absolute value at most $b_k$, so $|J_j|\le n_jb_k$ pointwise.
By \rlb{readk}, for $n_j>0$,
\eq
\Prob\bigl(|S_j-J_j|>t\mid\bssigma_{\Lambda_L\setminus\Lambda_L^j}\bigr)
\le2e^{-t^2/(2kn_j)}.\lb{conditional}
\en
Although the two conditionings differ, $|J_0|+|J_1|\le nb_k$ pointwise.
If $|n^{-1}\sum_{u\in\calU}W_u|>b_k+t$, at least one sublattice has $|S_j-J_j|>nt/2$.
Since $n_j\le n$, integration of \rlb{conditional} and a union bound yield
\eq
\mu_{L,\beta}^{\bc}\left(\left|\frac1n\sum_{u\in\calU}W_u\right|>b_k+t\right)
\le4\exp\left[-\frac{nt^2}{8k}\right].\lb{parityLD}
\en
Since $F_{\Gamma,L}^{\bc}=N^{-1}\sum_{u\in\calU}W_u$ and $n\le N$, the same bound holds with $|F_{\Gamma,L}^{\bc}|$ on the left.

Choose a fixed $K=K(d,\beta,\delta)$ such that $\rho^{K/2}\le\delta/4$.
For every $k\ge K$, the absolute value of each relevant thermal target is at most $\delta/4$ by \rlb{smallmean}.
Thus \rlb{parityLD}, with $t=\delta/2$, gives
\eq
\mu_{L,\beta}^{\bc}(|F_{\Gamma,L}^{\bc}-a_\Gamma^{\bc}(\beta)|>\delta)
\le4\exp\left[-\frac{\delta^2 n}{32k}\right],\qquad k\ge K.\lb{largeorder}
\en
For this estimate the target can be the unique-state, plus-state, or minus-state expectation, irrespective of the finite-volume boundary condition.
In particular, it can be used for the branches of a periodic cat.
The threshold $K$ may be large at low temperature but is independent of $L$.

For odd $L$ with periodic boundary conditions, the same proof uses a proper coloring with $2d+1$ colors instead of the bipartite partition, changing only the constants in \rlb{largeorder} and the fixed threshold $K$, which we henceforth choose uniformly for both parities of $L$.

\section{Simultaneous concentration, exact-energy conditioning, and typicality}
\label{sec:conditioning}
We assemble the bounded- and high-order estimates into a simultaneous canonical bound, then transfer it to the exact-energy ensemble used in the Letter. This proves the thermodynamic propositions and gives the typicality and linear-combination statements.

\label{sec:union}
Put $m=g(L)^d$. There are at most $C_K(1+m)^K$ subsets with $1\le|\Gamma|<K$, and at most $2^m$ nonempty subsets in total.
Combining \rlb{boundedorder} and \rlb{largeorder}, and using $n\ge(L-g(L)+1)^d\ge N/2$ for large $L$, gives
\eq
\mu_{L,\beta}^{\bc}(P_{\neqs}^{\bc})
\le C(1+m)^K e^{-cA_L^{\bc}}
+C\exp\left[(\log2)m-c'\frac Nm\right].\lb{unionfull}
\en
If $g(L)=o(\sqrt L)$, then $m^2=o(N)$.
The second exponent absorbs the factor $2^m$, while $\log(1+m)=O(\log N)$ is negligible compared with $A_L^{\bc}$.
This proves a canonical bound $e^{-c_0R_L^{\bc}}$, with $R_L^{\bc}$ as in \rlb{ratehigh} and \rlb{ratelow}, after changing $c_0$.
Notice that $R_L^{\bc}/\log N\to\infty$.

Here is a complete account of the exact-energy conditioning.
Let $E_*=E_L^{\bc}(\beta)$, $D=D_L^{\bc}(E_*)$, and $r=r_L^{\bc}$.
Here we introduce the projection $P_L^{\bc}$ onto $\calH_L^{\bc}(\beta)$ and the microcanonical density matrix
\eq
\Omega_L^{\bc}=\frac{P_L^{\bc}}{D},\qquad
\Tr(\Omega_L^{\bc}Q)=\bkt{Q}_{L,\beta}^{\mc,\bc}.\lb{SMOmega}
\en
Since $D e^{-\beta E_*}$ is the largest term in the partition sum,
\eq
Z_L^{\bc}(\beta)=\sum_E D_L^{\bc}(E)e^{-\beta E}
\le rD e^{-\beta E_*}.\lb{conditioningZ}
\en
For any positive operator $Q$,
\eq
\frac{\Tr(Qe^{-\beta H_L^{\bc}})}{Z_L^{\bc}(\beta)}
\ge\frac{e^{-\beta E_*}\Tr(P_L^{\bc}Q)}{Z_L^{\bc}(\beta)}
\ge\frac1r\Tr(\Omega_L^{\bc}Q).\lb{conditioningbound}
\en
The argument needs positivity but not $[Q,H_L^{\bc}]=0$.
Since $r=O(N)$, the canonical bound from \rlb{unionfull} proves Proposition~\ref{PRthermal}.
Applying the same argument to \rlb{magperiodic} proves the first part of Proposition~\ref{PRcat}.
The second part follows from spin flip, as explained in the End Matter.

For any shell projection $P$ of dimension $D$, set $\Omega=P/D$. For any positive $Q$, a Haar-random normalized $\ket\Phi\in\operatorname{Ran}P$ has
$\E\bra\Phi Q\ket\Phi=\Tr(PQ)/D$.
Thus, for example,
\eq
\Tr(\Omega Q)\le e^{-c_0 R}
\quad\Longrightarrow\quad
\Prob\{\bra\Phi Q\ket\Phi>e^{-c_0R/2}\}\le e^{-c_0R/2}.\lb{typicality}
\en
This proves typicality of MATE in the unperturbed energy shell.
For cats, one combines \rlb{typicality} for $P_{\out}$ with \rlb{EMHaar} for $B_L$ and its zero trace.

The perturbed and unperturbed shell projections have equal rank and obey
$\norm{\ttP_L^{\bc}-P_L^{\bc}}\le C\lambda_L$ by a resolvent contour around the isolated level.
For $\ttOmega_L^{\bc}=\ttP_L^{\bc}/D$, therefore,
\eq
\left|\Tr[(\ttOmega_L^{\bc}-\Omega_L^{\bc})Q]\right|
\le2\norm{\ttP_L^{\bc}-P_L^{\bc}}\norm Q
\le C\lambda_L\norm Q.\lb{perturbtyp}
\en
The first inequality uses $\operatorname{rank}(\ttP-P)\le2D$.
Shrinking $\lambda_L$ further preserves the exponentially small thermodynamic upper bounds, with slightly smaller constants, in the perturbed shell, and \rlb{typicality} then applies there too. The exact spin-flip trace identities are used in the unperturbed shell; the perturbation need not preserve them.

\subsection{Linear combinations of the macroscopic observables}
We make explicit the consequence of simultaneous measurement control for a real linear combination of the commuting observables. The exceptional probabilities are unchanged, while the density precision is multiplied by the sum of the absolute values of the coefficients.
For real coefficients $c_{\Gamma,L}$, let
\eq
A_L=\sum_\Gamma c_{\Gamma,L}O_{\Gamma,L}^{\bc},\qquad
a_{A,L}=\sum_\Gamma c_{\Gamma,L}a_\Gamma^{\bc}(\beta),\qquad
C_L=\sum_\Gamma|c_{\Gamma,L}|.\lb{linearcomb}
\en
On the common equilibrium subspace, every joint spectral value satisfies
$|A_L/N-a_{A,L}|\le\delta C_L$. The joint spectral calculus therefore gives
\eq
\Proj[|A_L/N-a_{A,L}|>\delta C_L]\le P_{\neqs}^{\bc}.\lb{linearproj}
\en
At every good time in Theorems~\ref{THhigh} and \ref{THlow}, the probability of violating this linear-combination condition is at most $\epsL$.
The same joint measurement event guarantees the condition for every real coefficient vector simultaneously, so no additional union bound or exceptional time set is needed.
Uniformly bounded $C_L$ permits any fixed desired precision after choosing the original $\delta$ accordingly.
A bound on $\norm{A_L/N}$ alone does not bound $C_L$ and is not the premise of this corollary.

\section{Thermal equilibrium in each cat branch}
\label{sec:branches}
We strengthen the magnetization-only cat statement by imposing all the appropriate spin-correlation conditions separately in the two phases. The goal is a measurement-based thermalization statement for each normalized branch, not a statement that its reduced density matrix is Gibbsian.

Fix $d\ge2$, $\beta>\betac(d)$, $0<\Delta<\ms$, $0<\delta<1$, and $g(L)=o(\sqrt L)$.
Use the periodic densities $F_{\Gamma,L}^{\pp}$ of \rlb{observables}, but define separate phase targets
\eq
 a_\Gamma^+=\mu_\beta^+(\sigma_\Gamma),\qquad
 a_\Gamma^-=(-1)^{|\Gamma|}a_\Gamma^+.\lb{branchtarget}
\en
Set
\eq
Q_\pm=1-\prod_{\varnothing\ne\Gamma\subset C_{g(L)}}
\Proj[|F_{\Gamma,L}^{\pp}-a_\Gamma^\pm|\le\delta],\qquad
\Pi_\pm=P_\pm(1-Q_\pm),\qquad
Q_{\rm br}=1-\Pi_+-\Pi_-.\lb{branchproj}
\en
These are commuting projections; $\Pi_+\Pi_-=0$ and $U_L\Pi_+U_L^*=\Pi_-$.

\para{Lemma S2}
For $R_L=\min\{L^{d-1},N/g(L)^d\}$ there is $c>0$ such that
\eq
\mu_{L,\beta}^{\pp}(Q_{\rm br})\le e^{-cR_L},\qquad
\Tr[\Omega_L^{\pp}(\Pi_+-\Pi_-)]=0.\lb{branchthermo}
\en
The first bound also holds microcanonically after reducing $c$.

\para{Proof}
Use the threshold $K$ of Sec.~\ref{sec:highorder}.
Choose a fixed $a>0$ small compared with $\Delta$, $\ms$, and $\delta/K$.
Couple the periodic configuration $\bssigma$ below the independent plus-block configuration $\bseta$ used in Sec.~\ref{sec:comparison}, with block side large enough that $\nu_L M\le\ms+a$.
For a product of order $k<K$, on the event $M(\bssigma)\ge\ms-a$ and $M(\bseta)\le\ms+2a$,
\eq
|F_{\Gamma,L}^{\pp}(\bseta)-F_{\Gamma,L}^{\pp}(\bssigma)|\le3ka.\lb{branchcouple}
\en
The mean under $\nu_L$ is uniformly close to the plus target. Indeed, coupling each plus block above the infinite-volume plus state and allowing for translates crossing the torus cut gives
\eq
|\nu_LF_{\Gamma,L}^{\pp}-a_\Gamma^+|
\le k(\nu_LM-\ms)+O_d(g(L)/L)\le ka+o(1).\lb{branchmean}
\en
This blockwise domination can be realized simultaneously because the product of plus-block measures dominates the restriction of the infinite-volume plus measure, by the same conditional-field comparison.
The independent-block tails for $M(\bseta)$ and $F_{\Gamma,L}^{\pp}(\bseta)$ have volume order.
Choosing, for example, $a<\delta/(16K)$ and taking $L$ large, \rlb{branchcouple} and \rlb{branchmean} imply
\eq
\mu_{L,\beta}^{\pp}
\bigl(M\ge\ms-a,\ |F_{\Gamma,L}^{\pp}-a_\Gamma^+|>\delta\bigr)
\le C e^{-cN},\qquad k<K.\lb{branchlow}
\en
Spin flip gives the corresponding negative-sector estimate.

For $k\ge K$, \rlb{largeorder} applies with either phase target without imposing any magnetization condition.
Except on the two-peak exceptional event at tolerance $a$, the system is in one of the two narrower magnetization windows, each contained in its corresponding $P_\pm$ window.
A union bound using \rlb{branchlow}, \rlb{largeorder}, and \rlb{magperiodic} therefore gives
\eq
\mu_{L,\beta}^{\pp}(Q_{\rm br})
\le Ce^{-cL^{d-1}}+C(1+g(L)^d)^K e^{-c'N}
+C2^{g(L)^d}e^{-c''N/g(L)^d}\le e^{-c'''R_L}.\lb{branchunion}
\en
Conditioning on the selected exact energy preserves the bound.
Spin flip preserves the exact-energy shell and exchanges $\Pi_+$ with $\Pi_-$, proving the exact zero trace in \rlb{branchthermo}.
\qedm

\para{Theorem S3 (thermal cat branches)}
For the preceding parameters there is $c>0$ such that, for sufficiently large $L$, one can choose a deterministic $\lambda_L>0$ with the following property.
With probability at least $1-\epsL$, where $\epsL=e^{-cR_L}$, every perturbed shell eigenstate is a balanced cat for the two projections $\Pi_\pm$ with error $\epsL^2/4$.
Every normalized pure initial state in the shell becomes such a cat with error $\epsL$ outside an atypical subset of relative length at most $\epsL$ in a sufficiently long interval $[0,T]$.

\para{Proof and interpretation}
Apply the proof of Theorem~\ref{THcat} with $P_{\out}$ replaced by $Q_{\rm br}$ and $B_L$ by $\Pi_+-\Pi_-$.
Lemma S2 supplies the two trace statements, and $R_L\le L^{d-1}=o(N)$ ensures that the Haar estimates with exponentially small errors remain valid.
All the remaining steps in the End Matter apply unchanged.

To describe the branches defined only by the magnetization measurement, let $\ket\Phi$ be a typical-time state given by this theorem and put $p_\pm=\bra\Phi P_\pm\ket\Phi$.
Since $P_\pm\ge\Pi_\pm$ and $\bra\Phi Q_{\rm br}\ket\Phi\le\epsL$, we have $p_\pm\ge1/2-\epsL$.
For the normalized branch $\ket{\Phi_\pm}=P_\pm\ket\Phi/\sqrt{p_\pm}$,
\eq
\bra{\Phi_\pm}Q_\pm\ket{\Phi_\pm}
=\frac{\bra\Phi P_\pm Q_\pm\ket\Phi}{p_\pm}
\le\frac{\epsL}{1/2-\epsL}\le3\epsL\lb{branchconditional}
\en
for large $L$.
Thus a simultaneous measurement of the whole family in a normalized branch gives its phase-specific thermal values with probability at least $1-3\epsL$.
The statement concerns the conditional macroscopic measurements within each pure branch, not a claim that its microscopic density matrix is Gibbsian.
\qedm

For $d=2$ the branch error scale is the same surface scale $L$ as in Theorem~\ref{THcat}.
For $d>2$, the magnetization-only cat theorem retains its surface scale even when the increasing correlation family yields the smaller scale $R_L$.
The perturbative good events in the End Matter can be chosen to hold throughout sufficiently small positive intervals of the perturbation strength. Taking the smaller of the two interval endpoints and intersecting the good events therefore makes both conclusions hold for the same perturbation, with their respective errors (after adjusting constants).

\section{Longer supports at fixed maximal order}
\label{sec:longer}
Here we distinguish the diameter of a product's support from the number of spins it contains. If the latter is bounded independently of $L$, we may enlarge the support box from $g(L)=o(\sqrt L)$ to $g(L)=o(L)$, while retaining volume-order errors in the disordered phase and surface-order errors in the ordered phase. We define this restricted family and state its thermalization and thermal-branch theorems explicitly.

\subsection{The restricted observable family}
We keep every product involving at most a fixed number of spins, without imposing a fixed bound on their separations. The energy shells, boundary conditions, and perturbation ensemble are unchanged from the Letter; only the family of observables is different.

Fix a positive integer $k_0$, independently of $L$, and a positive integer-valued function $g(L)$ satisfying $g(L)/L\to0$ as $L\uparrow\infty$.
In this section, no condition $g(L)=o(\sqrt L)$ is imposed.
Put $C_{g(L)}=\{0,\ldots,g(L)-1\}^d$ and define the set of allowed supports by
\eq
\calS_L^{(k_0)}=\{\Gamma\subset C_{g(L)}:1\le|\Gamma|\le k_0\}.\lb{fixedsupports}
\en
For each $\Gamma\in\calS_L^{(k_0)}$, define
\eq
O_{\Gamma,L}^{\bc}=\sum_{u\in\calU_{\Gamma,L}^{\bc}}\prod_{v\in\Gamma}Z_{u+v},\qquad
F_{\Gamma,L}^{\bc}=\frac{O_{\Gamma,L}^{\bc}}N,\qquad
\calO_L^{\bc,(k_0)}=\{O_{\Gamma,L}^{\bc}:\Gamma\in\calS_L^{(k_0)}\},\lb{fixedobservables}
\en
where $\calU_{\Gamma,L}^{\pp}=\Lambda_L$, with addition modulo $L$, and
$\calU_{\Gamma,L}^+=\{u\in\Zb^d:u+\Gamma\subset\Lambda_L\}$.
Thus the definitions of the observables and their normalization agree with \rlb{observables}, but the support box may now be much larger and the product order is bounded by $k_0$.
Every elementary Pauli product still has norm one.
For example, $k_0=2$ includes the magnetization and all two-spin correlation densities whose sites fit in a translate of $C_{g(L)}$, even when the distance between them diverges with $L$.
This is not the family of all products on the growing box: those of order greater than $k_0$ are excluded.

For thermalization, use $\bc=\pp$ when $0<\beta\le\betac(d)$ and $\bc=+$ when $\beta>\betac(d)$.
The target $a_\Gamma^{\bc}(\beta)$ is the expectation of $\prod_{v\in\Gamma}Z_v$ in the unique infinite-volume Gibbs state or the plus state, respectively, exactly as in the Letter.
For fixed $0<\delta<1$, define
\eq
P_{\neqs}^{\bc,(k_0)}
=1-\prod_{\Gamma\in\calS_L^{(k_0)}}
\Proj\bigl[|F_{\Gamma,L}^{\bc}-a_\Gamma^{\bc}(\beta)|\le\delta\bigr].\lb{fixedPneq}
\en
Its expectation in a normalized state is the probability that a simultaneous measurement of the entire family $\calO_L^{\bc,(k_0)}$ violates at least one thermal condition.
In particular, the singleton support gives $F_{\{0\},L}^{\bc}=M_L$ exactly, for both boundary conditions.

\subsection{Thermalization theorems}
The following statements are the fixed-maximal-order versions of Theorems~\ref{THhigh} and~\ref{THlow}. They retain the same temperature ranges, and the error exponents no longer decrease as the support box grows.

As in the Letter, draw $V_L=V_L^\dagger$ from normalized Lebesgue measure on the operator-norm unit ball in the full self-adjoint operator space, and set $\ttH_L^{\bc}=H_L^{\bc}+\lambda_LV_L$.
The shell $\ttcalH_L^{\bc}(\beta)$ is the spectral subspace for $(E_L^{\bc}(\beta)-2,E_L^{\bc}(\beta)+2)$, where $E_L^{\bc}(\beta)$ maximizes $D_L^{\bc}(E)e^{-\beta E}$.

\para{Theorem S4 (fixed-order thermalization in the disordered phase)}
Let $d\ge1$, $0<\beta\le\betac(d)$ with $\beta<\infty$, and fix $k_0$, $\delta$, and $g(L)=o(L)$ as above.
There is $c>0$ such that, for all sufficiently large $L$, one can choose an arbitrarily small deterministic $0<\lambda_L<1$ with the following property.
With probability at least $1-\epsL$, where $\epsL=e^{-cL^d}$, every normalized energy eigenstate $\ket\Psi\in\ttcalH_L^{\pp}(\beta)$ satisfies
\eq
\bra\Psi P_{\neqs}^{\pp,(k_0)}\ket\Psi\le\epsL^2/2.\lb{fixedETHhigh}
\en
For the same perturbation, every normalized initial state $\ket{\Phi(0)}\in\ttcalH_L^{\pp}(\beta)$ admits a sufficiently large $T$ and a measurable $\calA\subset[0,T]$ such that
\eq
\frac{|\calA|}{T}\le\epsL,\qquad
\bra{\Phi(t)}P_{\neqs}^{\pp,(k_0)}\ket{\Phi(t)}\le\epsL
\quad(t\in[0,T]\setminus\calA),\qquad
\ket{\Phi(t)}=e^{-i\ttH_L^{\pp}t}\ket{\Phi(0)}.\lb{fixeddynamicshigh}
\en

\para{Theorem S5 (fixed-order thermalization in the ordered phase)}
Let $d\ge2$, $\betac(d)<\beta<\infty$, and fix $k_0$, $\delta$, and $g(L)=o(L)$ as above.
There is $c>0$ such that, for all sufficiently large $L$, one can choose an arbitrarily small deterministic $0<\lambda_L<1$ with the following property.
With probability at least $1-\epsL$, where $\epsL=e^{-cL^{d-1}}$, every normalized energy eigenstate $\ket\Psi\in\ttcalH_L^+(\beta)$ satisfies
\eq
\bra\Psi P_{\neqs}^{+,(k_0)}\ket\Psi\le\epsL^2/2.\lb{fixedETHlow}
\en
For the same perturbation, every normalized initial state $\ket{\Phi(0)}\in\ttcalH_L^+(\beta)$ admits a sufficiently large $T$ and a measurable $\calA\subset[0,T]$ such that
\eq
\frac{|\calA|}{T}\le\epsL,\qquad
\bra{\Phi(t)}P_{\neqs}^{+,(k_0)}\ket{\Phi(t)}\le\epsL
\quad(t\in[0,T]\setminus\calA),\qquad
\ket{\Phi(t)}=e^{-i\ttH_L^+t}\ket{\Phi(0)}.\lb{fixeddynamicslow}
\en

In these theorems, $c$ may depend on $d,\beta,\delta,k_0$, but not on the initial state; the lower bound on $L$ also depends on the chosen function $g$.
The good event for $V_L$ is independent of the initial state, whereas $T$ and $\calA$ may depend on it.
At every good time, one simultaneous measurement of all the observables in $\calO_L^{\bc,(k_0)}$ therefore recovers their thermal densities within $\delta$, with probability at least $1-\epsL$.
In the ordered phase this includes a magnetization outcome within $\delta$ of $\ms(\beta)$.
As before, no lower bound on the admissible perturbation strength or upper bound on $T$ is obtained.

\subsection{Thermal cat branches for the restricted family}
We also impose the fixed-order thermal conditions separately on the two magnetization branches of a periodic cat. The magnetization-only statement of Theorem~\ref{THcat} needs no support restriction; the following result adds branch thermalization with $g(L)=o(L)$ while retaining its surface-order error scale.

Fix $d\ge2$, $\betac(d)<\beta<\infty$, $0<\Delta<\ms(\beta)$, and the preceding $k_0$, $\delta$, and $g$.
Use the periodic densities in \rlb{fixedobservables} and the phase targets
$a_\Gamma^+=\mu_\beta^+(\sigma_\Gamma)$ and $a_\Gamma^-=(-1)^{|\Gamma|}a_\Gamma^+$.
Keep the magnetization projections $P_\pm$ in \rlb{catproj}, and define
\eq
\begin{gathered}
Q_\pm^{(k_0)}=1-\prod_{\Gamma\in\calS_L^{(k_0)}}
\Proj[|F_{\Gamma,L}^{\pp}-a_\Gamma^\pm|\le\delta],\qquad
\Pi_\pm^{(k_0)}=P_\pm(1-Q_\pm^{(k_0)}),\\
Q_{\rm br}^{(k_0)}=1-\Pi_+^{(k_0)}-\Pi_-^{(k_0)},\qquad
B_{\rm br}^{(k_0)}=\Pi_+^{(k_0)}-\Pi_-^{(k_0)}.
\end{gathered}\lb{fixedbranchproj}
\en
The projection $\Pi_\pm^{(k_0)}$ imposes both the corresponding magnetization window and all the phase-specific thermal conditions.

\para{Theorem S6 (fixed-order thermal cat branches)}
For these parameters, there is $c>0$ such that, for all sufficiently large $L$, a deterministic $0<\lambda_L<1$ can be chosen with the following property.
With probability at least $1-\epsL$, where $\epsL=e^{-cL^{d-1}}$, every normalized energy eigenstate $\ket\Psi\in\ttcalH_L^{\pp}(\beta)$ satisfies
\eq
\bra\Psi Q_{\rm br}^{(k_0)}\ket\Psi\le\epsL^2/4,\qquad
\bigl|\bra\Psi B_{\rm br}^{(k_0)}\ket\Psi\bigr|\le\epsL^2/4.\lb{fixedbranchETH}
\en
For the same perturbation and every normalized pure initial state $\ket{\Phi(0)}\in\ttcalH_L^{\pp}(\beta)$, there are a sufficiently large $T$ and a measurable $\calA\subset[0,T]$ with $|\calA|/T\le\epsL$ such that
\eq
\bra{\Phi(t)}Q_{\rm br}^{(k_0)}\ket{\Phi(t)}\le\epsL,\qquad
\bigl|\bra{\Phi(t)}B_{\rm br}^{(k_0)}\ket{\Phi(t)}\bigr|\le\epsL
\quad(t\in[0,T]\setminus\calA),\lb{fixedbranchdynamics}
\en
where $\ket{\Phi(t)}=e^{-i\ttH_L^{\pp}t}\ket{\Phi(0)}$.
Thus these eigenstates and typical-time states are almost balanced pure cats for the refined projections $\Pi_\pm^{(k_0)}$.

For the branches selected by magnetization alone, write $p_\pm(t)=\bra{\Phi(t)}P_\pm\ket{\Phi(t)}$ and
$\ket{\Phi_\pm(t)}=P_\pm\ket{\Phi(t)}/\sqrt{p_\pm(t)}$.
At every good time, $p_\pm(t)\ge1/2-\epsL$, and the same calculation as in \rlb{branchconditional} gives
\eq
\bra{\Phi_\pm(t)}Q_\pm^{(k_0)}\ket{\Phi_\pm(t)}
\le\frac{\epsL}{1/2-\epsL}\le3\epsL\lb{fixedbranchconditional}
\en
for all sufficiently large $L$.
A simultaneous measurement in either normalized branch therefore satisfies every phase-specific condition for the restricted family with probability at least $1-3\epsL$.
This conclusion has the surface-order scale $L^{d-1}$ in every $d\ge2$, even for supports with $g(L)=o(L)$.

\subsection{Proofs of the fixed-order statements}
Only the bounded-order estimates are needed here. The number of supports grows polynomially in $L$, so the union bound does not reduce the volume- or surface-order concentration scale.

Indeed,
\eq
|\calS_L^{(k_0)}|
=\sum_{k=1}^{\min\{k_0,g(L)^d\}}\binom{g(L)^d}{k}
\le C_{k_0}(1+g(L)^d)^{k_0}.\lb{fixedcount}
\en
For $1\le|\Gamma|\le k_0$, the estimate \rlb{boundedorder} is already uniform in the support positions under the sole condition $g(L)=o(L)$.
Combining it with \rlb{fixedcount} gives
\eq
\mu_{L,\beta}^{\bc}(P_{\neqs}^{\bc,(k_0)})
\le C(1+g(L)^d)^{k_0}e^{-cA_L^{\bc}}
\le e^{-c'A_L^{\bc}},\qquad
A_L^{\bc}=\begin{cases}
N,&\beta\le\betac(d),\quad\bc=\pp,\\
L^{d-1},&\beta>\betac(d),\quad\bc=+.
\end{cases}\lb{fixedorderlong}
\en
Here the logarithm of the prefactor is $O(\log L)=o(A_L^{\bc})$.
The exact-energy comparison in Sec.~\ref{sec:conditioning} costs only another factor $O(N)$, which is absorbed in the same way.
The Fourier-basis and perturbative arguments in the End Matter then prove Theorems~S4 and~S5, with $R_L=A_L^{\bc}$ and a suitably smaller constant in $\epsL$.

For Theorem~S6, use the bounded-order part of the proof of Lemma~S2 for every $1\le|\Gamma|\le k_0$, choosing its auxiliary tolerance $a$ small enough for this fixed $k_0$.
The two-peak bound \rlb{magperiodic} controls the complement of the magnetization windows, and the phase-resolved product estimates have volume order and only polynomially many supports.
Consequently, after exact-energy conditioning and reducing $c_0>0$,
\eq
\bkt{Q_{\rm br}^{(k_0)}}_{L,\beta}^{\mc,\pp}\le e^{-c_0L^{d-1}},\qquad
\bkt{B_{\rm br}^{(k_0)}}_{L,\beta}^{\mc,\pp}=0.\lb{fixedbranchthermo}
\en
The second identity follows because spin flip preserves the shell and interchanges $\Pi_+^{(k_0)}$ and $\Pi_-^{(k_0)}$.
The proof of Theorem~\ref{THcat} in the End Matter, with $P_{\out}$ and $B_L$ replaced by $Q_{\rm br}^{(k_0)}$ and $B_{\rm br}^{(k_0)}$, now applies unchanged.
This proves Theorem~S6.\qedm

\clearpage
\section{Energy arithmetic and the strength of the perturbation}
\label{sec:arithmetic}
We verify the elementary spectral facts used to isolate the perturbed energy shell and to bound the cost of exact-energy conditioning. We also spell out why the random perturbation is generally nonlocal and many-body, despite its small total norm.

For $L\ge3$, the numbers of bonds, including fixed boundary bonds, are $dN$ for the torus and $d(N+L^{d-1})$ for plus boundaries.
In either geometry the all-plus configuration is a ground configuration with energy minus this bond count.
If $A\subset\Lambda_L$ is the set of flipped sites, its unhappy bonds form the edge boundary of $A$, including edges to fixed exterior sites in the plus case.
Let $b(A)$ count the interacting bonds with both endpoints in $A$, including wrapping bonds in the periodic case. Since every dynamical site has full degree $2d$,
\eq
|\partial A|=2d|A|-2b(A)\in2\Zb.\lb{cutparity}
\en
Each unhappy bond costs two, so
\eq
E_{\bssigma}-E_g=2|\partial A|\in4\Zb.\lb{energylattice}
\en
This proves the level-separation assertion used in the Letter.
It does not assert that every multiple of four occurs.
There are at most $dN+1$ distinct levels for the geometries considered here, since the total energy range is at most twice the bond count and the spacing is at least four.

The full-space distribution of $V_L$ is absolutely continuous and invariant under all unitary conjugations.
It is not restricted to finite-range, few-body, translation-invariant, or spin-flip-invariant operators.
In a Pauli expansion, coefficients of terms involving a macroscopic number of distant sites are generically nonzero.
Only its total norm is bounded, $\norm{\lambda_LV_L}\le\lambda_L$.
The deterministic choice of $\lambda_L$ in the End Matter depends on the desired errors and on the finite-size random spectral problem; no uniform positive lower bound follows.

\end{document}